\documentclass{article}

\usepackage{fancybox}
\usepackage{pstricks}
\usepackage{amsfonts}
\usepackage{graphicx}
\usepackage{algorithm}
\usepackage{algorithmic}
\usepackage{amsmath}
\usepackage{amssymb}
\usepackage[square,comma,sort]{natbib}
\usepackage{authblk}

\pdfoutput=1
\relax
\title{Accelerating Scientific Computations with Mixed Precision Algorithms}

\author[1,2]{Marc Baboulin}
\author[2]{Alfredo Buttari}
\author[2,3,4]{Jack Dongarra}
\author[2]{Jakub Kurzak}
\author[2]{Julie Langou}
\author[5]{Julien Langou}
\author[2]{Piotr Luszczek}
\author[2]{Stanimire Tomov}

\affil[1]{Department of Mathematics, University of Coimbra, Coimbra, Portugal}
\affil[2]{Department of Electrical Engineering and Computer Science, University Tennessee, Knoxville, Tennessee}
\affil[3]{Oak Ridge National Laboratory, Oak Ridge, Tennessee}
\affil[4]{University of Manchester, Manchester, UK}
\affil[5]{Department of Mathematical and Statistical Sciences, University of Colorado Denver, Denver, Colorado}

\begin{document}
\maketitle
\begin{abstract}

On modern architectures, the performance of 32-bit operations is often at least
twice as fast as the performance of 64-bit operations.  By using a combination
of 32-bit and 64-bit floating point arithmetic, the performance of many dense
and sparse linear algebra algorithms can be significantly enhanced while
maintaining the 64-bit accuracy of the resulting solution.  The approach
presented here can apply not only to conventional processors but also to
other technologies such as Field Programmable Gate Arrays (FPGA), Graphical
Processing Units (GPU), and the STI Cell BE processor.  Results on modern
processor architectures and the STI Cell BE are presented.

On modern architectures, the performance of 32-bit operations is often at least
twice as fast as the performance of 64-bit operations.  By using a combination
of 32-bit and 64-bit floating point arithmetic, the performance of many dense
and sparse linear algebra algorithms can be significantly enhanced while
maintaining the 64-bit accuracy of the resulting solution.  The approach
presented here can apply not only to conventional processors but also to
other technologies such as Field Programmable Gate Arrays (FPGA), Graphical
Processing Units (GPU), and the STI Cell BE processor.  Results on modern
processor architectures and the STI Cell BE are presented.

\end{abstract}

\noindent

\section{Introduction}
\label{sec:intro}

On modern architectures, the performance of 32-bit operations is often at least
twice as fast as the performance of 64-bit operations. There are two reasons
for this.
Firstly, 32-bit floating point arithmetic is usually twice as fast as 64-bit
floating point arithmetic on most modern processors.
Secondly the amount of bytes moved through the memory system is halved. In
Table~\ref{tab:procs_features}, we provide some hardware numbers that support
these claims.  On AMD Opteron 246, IBM PowerPC 970, and Intel Xeon 5100, the
single precision peak is twice the double precision peak. On the STI Cell BE,
the single precision peak is fourteen times the double precision peak.  Not
only single precision is faster than double precision on conventional
processors but this is also the case on less mainstream technologies such as Field
Programmable Gate Arrays (FPGA) and Graphical Processing Units (GPU).  These
speedup numbers tempt us and we would like to be able to benefit from it.

For several physics applications, results with 32-bit accuracy are not an option
and one really needs 64-bit accuracy maintained throughout the computations.
The obvious reason is for the application to give an accurate
answer. Also, 64-bit accuracy enables most of the modern computational
methods to be more stable; therefore, in critical conditions, one must use
64-bit accuracy to obtain an answer. In this manuscript, we present a
methodology of how to perform the bulk of the operations in 32-bit arithmetic, then
postprocess the 32-bit solution by refining it into a a solution that is 64-bit
accurate. We present this methodology in the context of solving a system of linear
equations, be it sparse or dense, symmetric positive definite or nonsymmetric,
using either direct
or iterative methods. We believe that the approach outlined below is
quite general and should be considered by application developers for their
practical problems.

\section{The Idea Behind Mixed Precision Algorithms}
\label{sec:itref}

Mixed precision algorithms stem from the observation that, in many cases, a
single precision solution of a problem can be refined to the point where double
precision accuracy is achieved.  The refinement can be accomplished, for
instance, by means of the Newton's algorithm~\cite{ypma:531} which computes the
zero of a function $f(x)$ according to the iterative formula
\begin{equation}
  \label{eq:newton}
  x_{n+1}=x_n-\frac{f(x_n)}{f'(x_n)}.
\end{equation}
In general, we would compute a starting point and $f'(x)$ in single precision
arithmetic and the refinement process will be computed in double precision
arithmetic.

If the refinement process is cheaper than the initial computation of the solution then
double precision accuracy can be achieved nearly at the same speed as the single
precision accuracy. Sections~\ref{sec:direct} and~\ref{sec:iterative}
describe how this concept can be applied to solvers of linear systems based on
direct and iterative methods, respectively.

\subsection{Direct Methods}
\label{sec:direct}

A common approach to the solution of linear systems, either dense or sparse, is
to perform the LU factorization of the coefficient matrix using Gaussian
elimination.  First, the coefficient matrix~$A$ is factored into the product of
a lower triangular matrix $L$ and an upper triangular matrix $U$.  Partial row
pivoting is in general used to improve numerical stability resulting in a
factorization $PA = LU$, where $P$ is a permutation matrix.  The solution for
the system is achieved by first solving $Ly = Pb$ ({\em forward substitution})
and then solving $Ux = y$ ({\em backward substitution}).  Due to round-off
errors, the computed solution $x$ carries a numerical error magnified by the
condition number of the coefficient matrix~$A$.

In order to improve the computed solution, we can apply an iterative process
which produces a correction to the computed solution at each iteration, which
then yields the method that is commonly known as the {\it iterative refinement}
algorithm.
As Demmel points out~\cite{Demmel_1997}, the non-linearity of the
round-off errors makes the iterative refinement process equivalent to the
Newton's method applied to the function $f(x) = b - Ax$.  Provided that the
system is not too ill-conditioned, the algorithm produces a solution correct to
the working precision.  
Iterative refinement in double/double precision is a
fairly well understood concept and was analyzed by
Wilkinson~\cite{Wilkinson_1963}, Moler~\cite{Moler_1967_jacm} and
Stewart~\cite{Stewart_1973}.

\begin{algorithm}[!h]
\caption{\label{alg:dir_itref}Mixed precision, Iterative Refinement
  for Direct Solvers}
\begin{tabbing}
  \footnotesize{1:} \hspace{1mm}\= $LU$\=$\leftarrow PA$  \hspace{18mm}\=\textcolor{blue}{($\varepsilon_s$)}\\
  \footnotesize{2:} \> solve $Ly=Pb$ \>\>\textcolor{blue}{($\varepsilon_s$)}\\
  \footnotesize{3:} \> solve $Ux_0=y$ \>\>\textcolor{blue}{($\varepsilon_s$)}\\
                    \> {\bf do}  $k=1, 2, ...$\\
  \footnotesize{4:} \>\>$r_k \leftarrow b-Ax_{k-1}$ \>\textcolor{blue}{($\varepsilon_d$)}\\
  \footnotesize{5:} \>\>solve $Ly=Pr_k$ \>\textcolor{blue}{($\varepsilon_s$)}\\
  \footnotesize{6:} \>\>solve $Uz_k=y$ \>\textcolor{blue}{($\varepsilon_s$)}\\
  \footnotesize{7:} \>\>$x_k \leftarrow x_{k-1}+z_k$ \>\textcolor{blue}{($\varepsilon_d$)}\\
  \>\> {\bf check convergence} \\
                    \> {\bf done}
\end{tabbing}
\end{algorithm}

The algorithm can be modified to use a mixed precision approach.  The
factorization $PA = LU$ and the solution of the triangular systems $Ly = Pb$
and $Ux = y$ are computed using single precision arithmetic.  The residual
calculation and the update of the solution are computed using double precision
arithmetic and the original double precision coefficients (see
Algorithm~\ref{alg:dir_itref}).  The most computationally expensive operation,
the factorization of the coefficient matrix $A$, is performed using single
precision arithmetic and takes advantage of its higher speed.  The only
operations that must be executed in double precision are the residual
calculation and the update of the solution (they are denoted with an
$\varepsilon_d$ in Algorithm~\ref{alg:dir_itref}).  We observe that the only
operation with computational complexity of $\mathcal{O}(n^3)$ is handled in single
precision, while all operations performed in double precision are of at most
$\mathcal{O}(n^2)$ complexity.  The coefficient matrix~$A$ is converted to
single precision for the LU factorization and the resulting factors are stored
in single precision while the initial coefficient matrix~$A$ needs to be kept
in memory. Therefore, one drawback of the following approach is that the it
uses 50\% more memory than the standard double precision algorithm.

The method in Algorithm~\ref{alg:dir_itref} can offer significant
improvements for the solution of a sparse linear system in many cases
if:
\begin{enumerate}

\item single precision computation is significantly faster than double
precision computation.

\item the iterative refinement procedure converges in a small number of steps.

\item the cost of each iteration is small compared to the cost of the system
factorization. If the cost of each iteration is too high, then a low number of
iterations will result in a performance loss with respect to the full double
precision solver. In the sparse case, for a fixed matrix size, both the cost of
the system factorization and the cost of the iterative refinement step may
substantially vary depending on the number of non-zeroes and the matrix sparsity
structure. In the dense case, results are more predictable.

\end{enumerate}

Note that the choice of the stopping criterion in the iterative refinement process
is critical. Formulas for the error computed at each step of 
Algorithm~\ref{alg:dir_itref} can be obtained for instance in~\cite{Oettli_1964,Demmel_2006}.

\subsection{Iterative Methods}
\label{sec:iterative}

Direct methods are usually a very robust tool for the solution of sparse
linear systems. However, they suffer from fill-in which results in high
memory requirements, long execution time, and non-optimal
scalability in parallel environments.  To overcome these limitations,
various pivot reordering techniques are commonly applied to minimize the amount of
generated fill-in and to enable better exploitation of parallelism.
Still, there are cases where direct methods pose too high of a memory
requirement or deliver poor performance.  A valid alternative are iterative
methods even though they are less robust and have a less
predictable behavior. Iterative methods do not require more memory than what is needed for
the original coefficient matrix. Further more, time to solution can be better
than that of direct methods if convergence is achieved in relatively few
iterations~\cite{templates,saad_book}.

In the context of iterative methods, the refinement method outlined
in Algorithm~\ref{alg:dir_itref} can be represented as
\begin{equation}\label{richardson}
          x_{i+1} = x_i + M (b - Ax_i),
\end{equation}
where~$M$ is $(LU)^{-1}P$. Iterative methods of this form (i.e. where~$M$ does
not depend on the iteration number~$i$) are also known as {\it stationary}.
Matrix~$M$ can be as simple as a scalar value~(the method then becomes a
modified Richardson iteration) or as complex as~$(LU)^{-1}P$. In either case,
$M$ is called a {\it preconditioner}. The preconditioner should approximate
$A^{-1}$, and the quality of the approximation determines the convergence
properties of~(\ref{richardson}). In general, a preconditioner is intended to
improve the robustness and the efficiency of iterative methods.  Note
that~(\ref{richardson}) can also be interpreted as a Richardson method's
iteration in solving $M A x = M b$ which is called {\sl left} preconditioning.
An alternative is to use {\it right} preconditioning, whereby the original
problem $Ax = b$ is transformed into a problem of solving
$$
              A M u = b, ~~x = M u
$$
iteratively.  Later on, we will use the right preconditioning for mixed
precision iterative methods.

$M$ needs to be easy to compute, apply, and store to guarantee
the overall efficiency. Note that these requirements were
addressed in the mixed precision direct methods above by replacing~$M$
(coming from LU factorization of $A$ followed by matrix inversion),
with its single precision representation so that arithmetic operations
can be performed more efficiently on it.
Here however, we go two steps further. We replace not
only~$M$ by an inner loop which is an incomplete iterative solver working in
single precision arithmetic~\cite{Turner92}. Also, the outer loop
is replaced by a more sophisticated iterative method~e.g., based on
Krylov subspace.

Note that replacing~$M$ by an iterative method leads to {\it nesting}
of two iterative methods. Variations of this type of nesting, also
known in the literature as an {\it inner-outer} iteration, have been
studied, both theoretically and
computationally~\cite{golub00inexact,saad91flexible,flexible-inner-outer,cg-vassilevski,notay00flexible,vuik-variable-GMRES,essg:03}.
The general appeal of these methods is that the computational speedup hinges
inner solver's ability to use an approximation of the original
matrix~$A$ that is fast to apply. In our case, we use
single precision arithmetic matrix-vector product as a fast approximation
of the double precision operator in the inner iterative solver.  Moreover, even if no faster
matrix-vector product is available, speedup can often be observed due
to improved convergence~(e.g., see~\cite{flexible-inner-outer}, where
Simoncini and Szyld explain the possible benefits of FGMRES-GMRES over
restarted GMRES).

To illustrate the above concepts, we demonstrate an inner-outer nonsymmetric
iterative solver in mixed precision. The solver is based on the restarted
Generalized Minimal RESidual~(GMRES) method.  In particular, consider
Algorithm~\ref{alg:pgmres}, where the outer loop uses the flexible
GMRES~(FGMRES~\cite{saad91flexible,saad_book}) and the inner loop uses the GMRES
in single precision arithmetic~(denoted by GMRES$_{SP}$). FGMRES, being a minor
modification of the standard GMRES, is meant to accommodate non-constant
preconditioners. Note that in our case, this non-constant preconditioner is
GMRES$_{SP}$. The resulting method is denoted by
FGMRES($m_{out}$)-GMRES$_{SP}$($m_{in}$) where~$m_{in}$ is the restart for the
inner loop and~$m_{out}$ for the outer FGMRES.  Algorithm~\ref{alg:pgmres}
checks for convergence every $m_{out}$ outer iterations.  Our actual
implementation checks for convergence at every inner iteration, this can be done
with simple tricks at almost no computational cost.

\begin{algorithm}[!h]
\caption{\label{alg:pgmres}~~ Mixed precision, inner-outer
        FGMRES($m_{out}$)-GMRES$_{SP}$($m_{in}$)}
\begin{algorithmic}[1]
  \FOR{$i=0,1,...$}
    \STATE $r=b-Ax_i$ \hfill\textcolor{blue}{($\varepsilon_d$)}
    \STATE $\beta = h_{1,0} = ||r||_2 $  \hfill\textcolor{blue}{($\varepsilon_d$)}
    \STATE check ~convergence~ and~ exit~ if~ done
    \FOR{$k=1, \dots, m_{out}$}
      \STATE $v_k = r ~/ ~h_{k, k-1}$  \hfill\textcolor{blue}{($\varepsilon_d$)}
      \STATE Perform one cycle of GMRES$_{SP}(m_{in})$ in order to (approximately) solve $A z_k = v_k$,
             (initial guess $z_k = 0$)  \hfill\textcolor{blue}{($\varepsilon_s$)}
      \STATE $r = A~z_k$  \hfill\textcolor{blue}{($\varepsilon_d$)}
      \FOR{j=1,\dots,k}
        \STATE  $h_{j,k} = r^T v_j$  \hfill\textcolor{blue}{($\varepsilon_d$)}
        \STATE  $r = r - h_{j,k}~ v_j$  \hfill\textcolor{blue}{($\varepsilon_d$)}
      \ENDFOR
      \STATE $h_{k+1,k} = || r ||_2$  \hfill\textcolor{blue}{($\varepsilon_d$)}
    \ENDFOR
    \STATE Define~ $Z_k~ = ~[z_1, \dots, z_k],~ H_k = \{h_{i,j} \}_{1 \le i \le k+1, 1 \le j \le k}$  \hfill\textcolor{blue}{($\varepsilon_d$)}
    \STATE Find ~ $y_k $, the vector of size $k$,  ~that ~minimizes~ $||\beta e_1-H_k ~y_k||_2$  \hfill\textcolor{blue}{($\varepsilon_d$)}
    \STATE $x_{i+1} = x_i + Z_k~ y_k$  \hfill\textcolor{blue}{($\varepsilon_d$)}
  \ENDFOR
\end{algorithmic}
\end{algorithm}

The potential benefits of FGMRES compared to GMRES are becoming better
understood~\cite{flexible-inner-outer}. Numerical experiments confirm
improvements in speed, robustness, and sometimes memory requirements for these
methods. For example, we show a maximum speedup of close to~$15$ on the
selected test problems.  The memory requirements for the method are the
matrix~$A$ in CRS format, the nonzero matrix coefficients in single precision,
$2~ m_{out}$ number of vectors in double precision, and~$m_{in}$ number of
vectors in single precision.

The Generalized Conjugate Residuals (GCR)
method~\cite{vuik-variable-GMRES,vandervorst91gmresr} is a possible
replacement for FGMRES as the outer iterative solver. Whether to choose
GCR or FGMRES is not yet well understood.

As in the dense case, the choice of the stopping criterion in the iterative
refinement process is critical. In the sparse case, formulas for the errors can
be computed following the work of Arioli et al.~\cite{Arioli_1989}.

\section{Performance Results}
\label{sec:perf}
The experimental results reported in this section were measured on the
systems described in Table~\ref{tab:procs_features}. At this moment no
software libraries are available to perform sparse computations on the
STI Cell BE architecture. For this reason, only mixed precision iterative
refinement solvers for dense linear systems are presented for this architecture.

\begin{table}[!h]
\caption{Hardware and software details of the systems used for
  performance experiments.}
\label{tab:procs_features}
\begin{tabular*}{\textwidth}{lccccc}
\hline
\textbf{Architecture} & \textbf{Clock} &  Peak SP        & Memory             & BLAS        & Compiler    \\
                      & \textbf{[GHz]} &  / Peak DP      & \textbf{[MB]}      &             &             \\
\hline
AMD Opteron 246       & 2.0            & 2               & 2048               & Goto-1.13   & Intel-9.1   \\
IBM PowerPC 970       & 2.5            & 2               & 2048               & Goto-1.13   & IBM-8.1     \\
Intel Xeon 5100       & 3.0            & 2               & 4096               & Goto-1.13   & Intel-9.1   \\
STI Cell BE           & 3.2            & 14              & 512                &   --        & Cell SDK-1.1\\
\hline
\end{tabular*}
\end{table}

To measure the performance of sparse mixed precision solvers based on
both direct and iterative methods, the matrices described in
Table~\ref{tab:matrices} were used.

\begin{table}[!h]
  \caption{Test matrices for sparse mixed precision, iterative refinement solution methods.}
  \label{tab:matrices}
  \begin{tabular*}{\textwidth}{llccccc}
    \hline
\textbf{n.} &    \textbf{Matrix}     &  \textbf{Size}   &
\textbf{Nonzeroes}&  \textbf{symm.} & \textbf{pos. def.} &\textbf{C. Num.} \\
    \hline
1    &    SiO        &  33401  &  1317655  &  yes   &  no    & $O(10^3)$ \\
2    &    Lin        &  25600  &  1766400  &  yes   &  no    & $O(10^5)$ \\
3    &    c-71       &  76638  &  859554   &  yes   &  no    & $O(10)$ \\
4    &    cage-11    &  39082  &  559722   &  no    &  no    & $O(1)$ \\
5    &    raefsky3   &  21200  &  1488768  &  no    &  no    & $O(10)$ \\
6    &    poisson3Db &  85623  &  2374949  &  no    &  no    & $O(10^3)$ \\
    \hline
\end{tabular*}
\end{table}

Based on backward stability analysis, the solution~$x$ can be
considered as accurate as the double precision one when
\begin{displaymath}
  \|b-Ax\|_2 \le \|x\|_2 \cdot \|A\|_2 \cdot \varepsilon \cdot \sqrt{n}
\end{displaymath}
where $\| \cdot \|_{2}$ is the spectral norm. However, for the
following experiments, a full double precision solution is computed
first and then the mixed precision iterative refinement is stopped
when the computed solution is as accurate as the full double precision one.

\subsection{Direct Methods}
\label{sec:perf_direct}

\subsubsection{Dense Matrices}
Mixed precision iterative refinement solvers were developed for both symmetric
and nonsymmetric dense linear systems by means of the methods and subroutines
provided by the BLAS~ \cite{blas01,blas02,blas03,blas04,blas05} and
LAPACK~\cite{lapack:99} software packages. For the nonsymmetric case, step 1 in
Algorithm~\ref{alg:dir_itref} is implemented by means of the \texttt{SGETRF}
subroutine, steps 2,3 and 5,6 with the \texttt{SGETRS} subroutine, step 4 with
the \texttt{DGEMM} subroutine and step 7 with the \texttt{DAXPY} subroutine. For the
symmetric case the \texttt{SGETRF}, \texttt{SGETRS} and \texttt{DGEMM}
subroutines were replaced by the \texttt{SPOTRF}, \texttt{SPOTRS} and
\texttt{DSYMM} subroutines, respectively. Further details on these implementations can
be found in~\cite{Langou_2006_sc,1297653}.

As already mentioned, iterative refinement solvers require $1.5$ times
as much memory as a regular double precision solver. It is because
the mixed precision iterative refinement solvers need to store at the
same time both the single precision and the double precision versions of the
coefficient matrix. It is true for dense as well as sparse matrices.

\begin{table}[!h]
  \centering
  \begin{tabular}{|r|cc|}
             \hline
                    &   Nonsymmetric  &  Symmetric \\
             \hline
AMD Opteron 246     &  1.82     & 1.54     \\
IBM PowerPC 970     &  1.56     & 1.35     \\
Intel Xeon 5100     &  1.56     & 1.43     \\
STI Cell BE         &  8.62     & 10.64 \\
\hline
  \end{tabular}
  \caption{Performance improvements for direct dense methods
when going from a full double precision solve (reference time) to a
mixed precision solve.}
  \label{tab:dir_dense}
\end{table}

\begin{figure*}
\begin{minipage}[tl]{0.5\textwidth}
\begin{center}
\includegraphics[width=\textwidth]{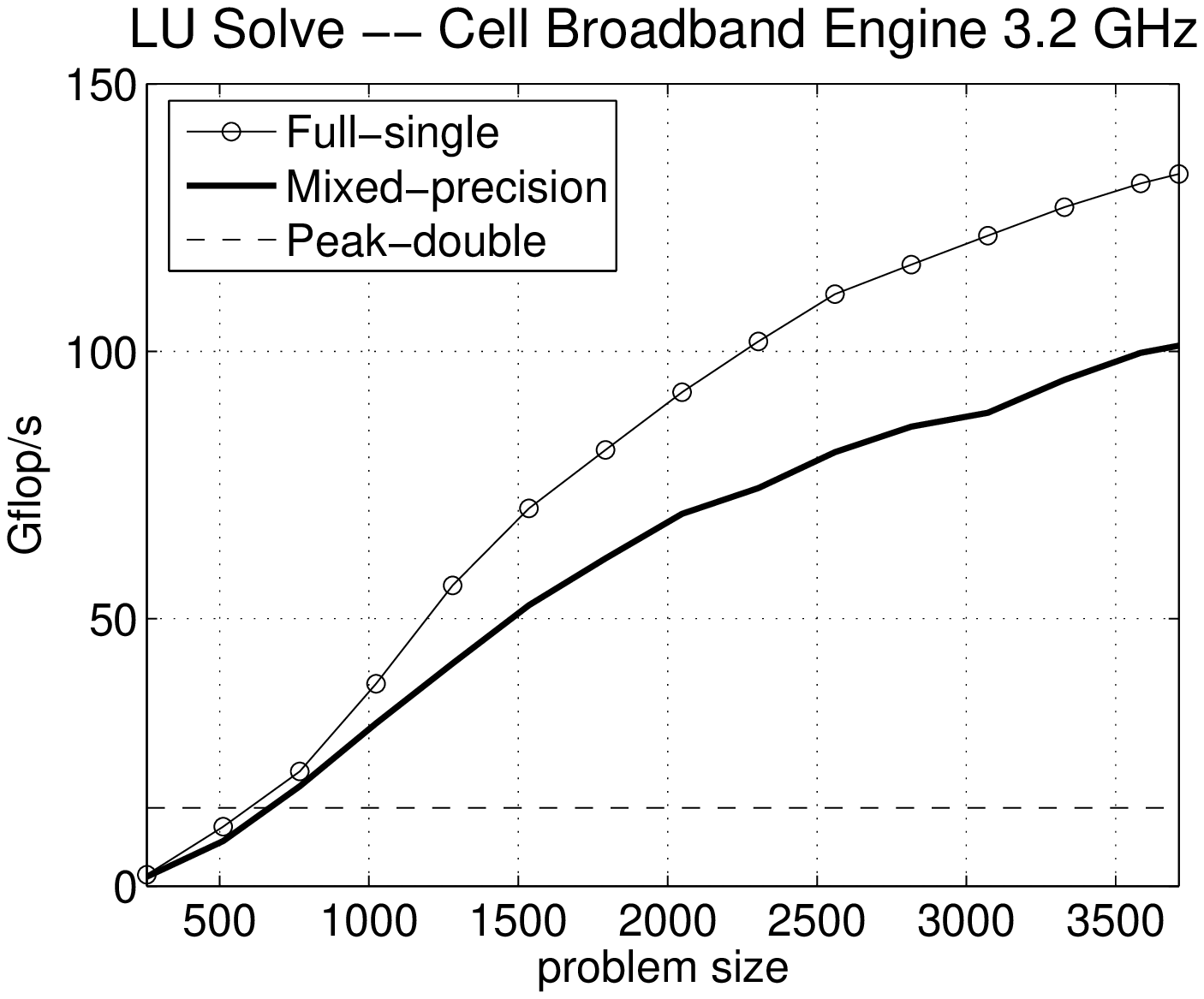}
\end{center}
\end{minipage}
\hspace{0.25cm}
\begin{minipage}[tr]{0.5\textwidth}
\begin{center}
\includegraphics[width=\textwidth]{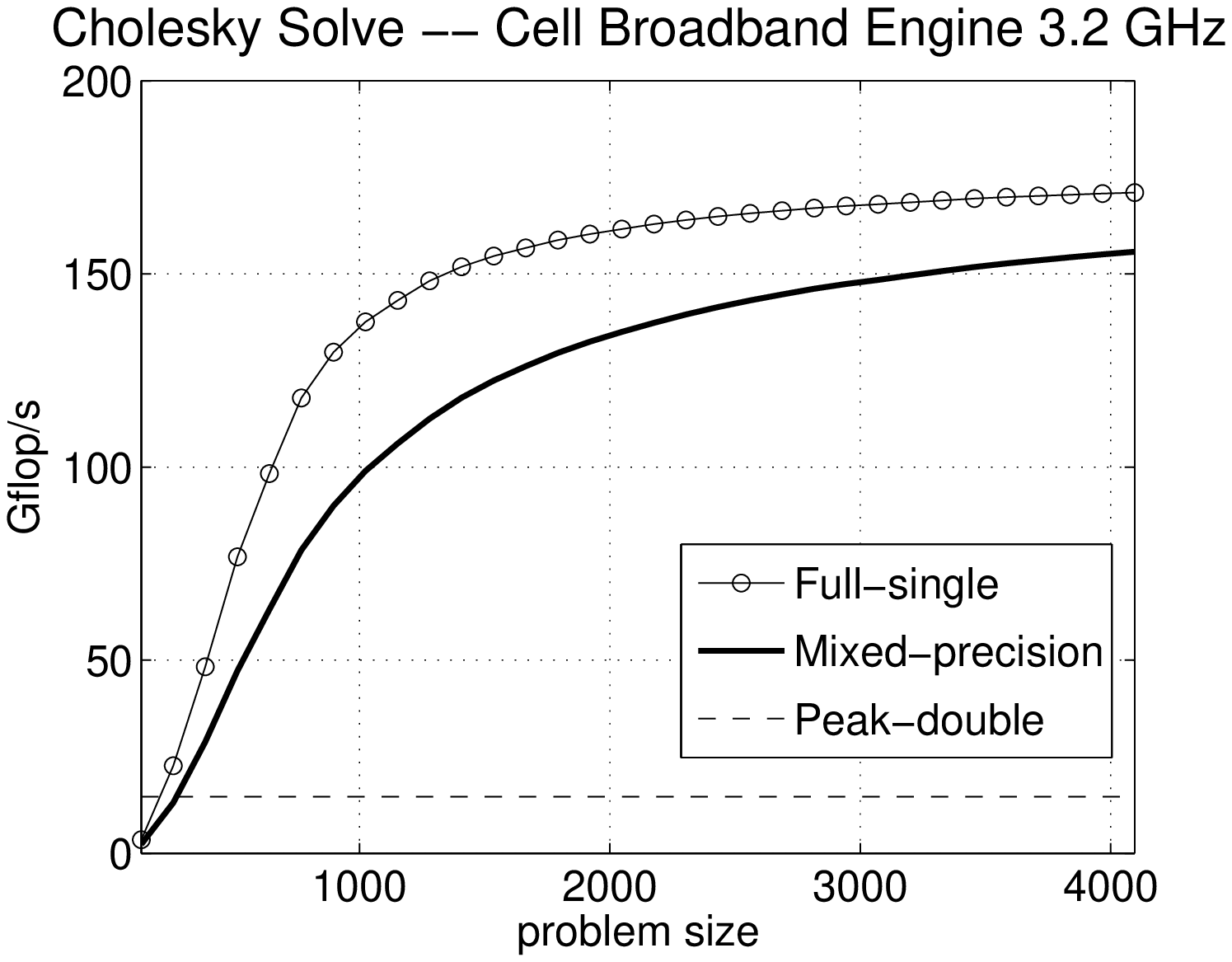}
\end{center}
\end{minipage}
\caption{\label{fig:cell} Mixed precision, iterative refinement method
  for the solution of dense linear systems on the STI Cell BE processor.}
\end{figure*}

Table~\ref{tab:dir_dense} shows the speedup of the mixed precision,
iterative refinement solvers for dense matrices with respect to full,
double precision solvers.
These results show that the mixed
precision iterative refinement method can run very close to the speed
of the full single precision solver while delivering the same accuracy
as the full double precision one. On the AMD Opteron, Intel Woodcrest
and IBM PowerPC architectures, the mixed precision, iterative solver
can provide a speedup of up to 1.8 for the nonsymmetric solver and 1.5
for the symmetric one for large enough problem sizes. For small
problem sizes the cost of even a few iterative refinement
iterations is high compared to the cost of the factorization and
thus the mixed precision iterative solver is less efficient than the
double precision one.

Parallel implementations of Algorithm~\ref{alg:dir_itref} for the symmetric and
nonsymmetric cases have been produced in order to exploit the full computational
power of the Cell processor (see also Figure~\ref{fig:cell}).
Due to the large difference between the speed of single
precision and double precision floating point units\footnote{As indicated in
Table~\ref{tab:procs_features}, the peak for single precision operations is 14
times more than the peak for double precision operations on the STI Cell BE.},
the mixed precision solver performs up to $7$ times faster than the double
precision peak in the nonsymmetric case and $11$ times faster for the symmetric
positive definite case. Implementation details for this case can be found
in~\cite{Kurzak_2007_cell_linpack,Kurzak_2007_cell_cholesky}.

\subsubsection{Sparse Matrices}

Most sparse direct methods for solving linear systems of equations are variants
of either multifrontal~\cite{duff1983} or supernodal~\cite{ashcraft1987}
factorization approaches. Here, we focus only on multifrontal methods. For
results on supernodal solvers see~\cite{sparse_iterref}.  There are a number of
freely available packages that implement multifrontal methods.  We have chosen
for our tests a software package called
MUMPS~\cite{amestoy2000,amestoy2001,amestoy2006}. The main reason for selecting
this software is that it is implemented in both single and double precision,
which is not the case for other freely available multifrontal solvers such as
UMFPACK~\cite{davis1997,davis1999,davis2004}.

Using the MUMPS package for solving systems of linear equations comprises
of three separate steps:
\begin{enumerate}
\item System Analysis: in this phase the system sparsity structure is
  analyzed in order to estimate the element fill-in, which provides an
  estimate of the memory that will be allocated in the following
  steps. Also, pivoting is performed based on the structure of $A+A^T$,
  ignoring numerical values. Only integer operations are performed at
  this step.
\item Matrix Factorization: in this phase the $PA=LU$ factorization is
  performed. This is the computationally most expensive step of the
   system solution.
\item System Solution: the system is solved in two steps: $Ly=Pb$ and
$Ux=y$.
\end{enumerate}

The Analysis and Factorization phases correspond to step 1 in
Algorithm~\ref{alg:dir_itref} while the solution phase correspond to steps 2,3
and 5,6.

\begin{table}[!h]
  \centering
  \begin{tabular}{|r|cccccc|}
\cline{2-7}
\multicolumn{1}{}{} & \multicolumn{6}{|c|}{Matrix number} \\
\cline{2-7}
\multicolumn{1}{c|}{} & 1     &  2     &  3     &  4     &  5     &  6     \\
\hline
AMD Opteron 246     & 1.827 &  1.783 &  1.580 &  1.858 &  1.846 &  1.611 \\
IBM PowerPC 970     & 1.393 &  1.321 &  1.217 &  1.859 &  1.801 &  1.463 \\
Intel Xeon 5100     & 1.799 &  1.630 &  1.554 &  1.768 &  1.728 &  1.524 \\
\hline
  \end{tabular}
  \caption{Performance improvements for direct sparse methods
when going from a full double precision solve (reference time) to a
mixed precision solve.}
  \label{tab:mumps}
\end{table}

\begin{figure}[!h]
\centering
  \includegraphics[width=0.8\textwidth]{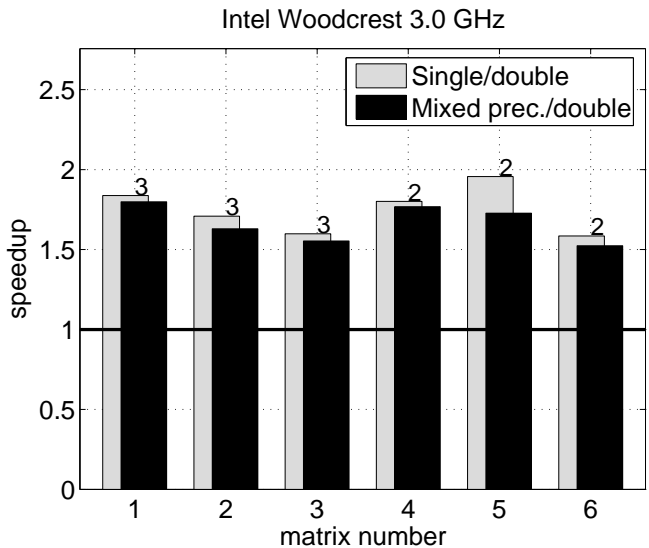}
  \caption{Mixed precision, iterative refinement
    with the MUMPS direct solver on an Intel Woodcrest 3.0 GHz system.}
  \label{fig:mumps_wood}
\end{figure}

The speedup of the mixed precision, iterative refinement approach over the
double precision one for sparse direct methods is shown in
Table~\ref{tab:mumps}, and Figure~\ref{fig:mumps_wood}. The figure reports the
performance ratio between the full single precision and full double precision
solvers~(light colored bars) and the mixed precision and full-double precision
solvers~(dark colored bars) for six matrices from real world applications. The
number on top of each bar shows how many iterations are performed by the mixed
precision, iterative method to achieve double precision accuracy.

\subsection{Iterative Methods}
\label{sec:perf_iterative}


Similar to the case of sparse direct solvers, we demonstrate the numerical
performance of Algorithm~\ref{alg:pgmres} on the architectures from
Table~\ref{tab:procs_features} and on the matrices from
Table~\ref{tab:matrices}.

\begin{figure*}
  \begin{minipage}[tl]{0.5\textwidth}
    \begin{center}
      \includegraphics[width=\textwidth]{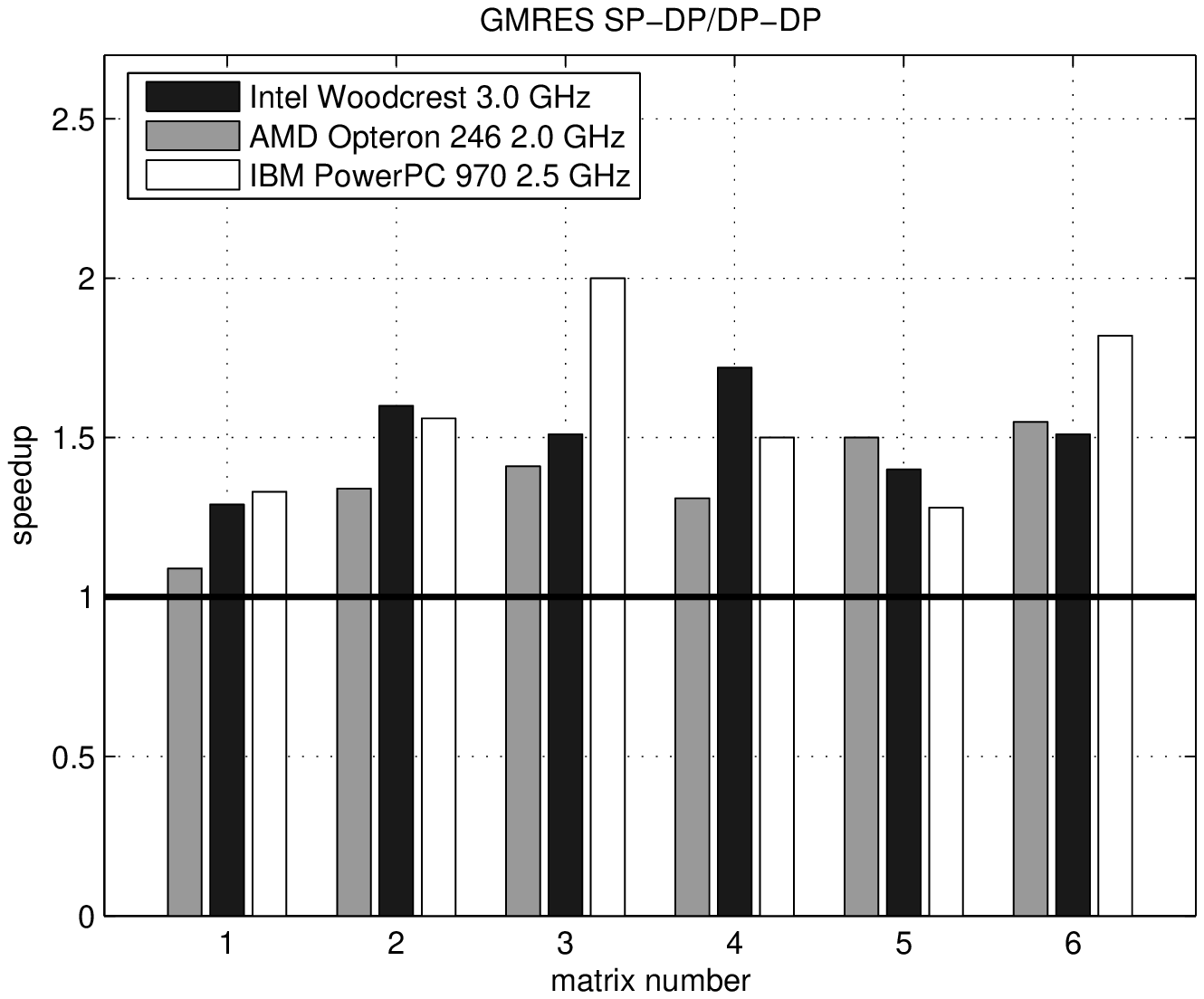}
    \end{center}
  \end{minipage}
  \hspace{0.25cm}
  \begin{minipage}[tr]{0.417\textwidth}
    \begin{center}
      \includegraphics[width=\textwidth]{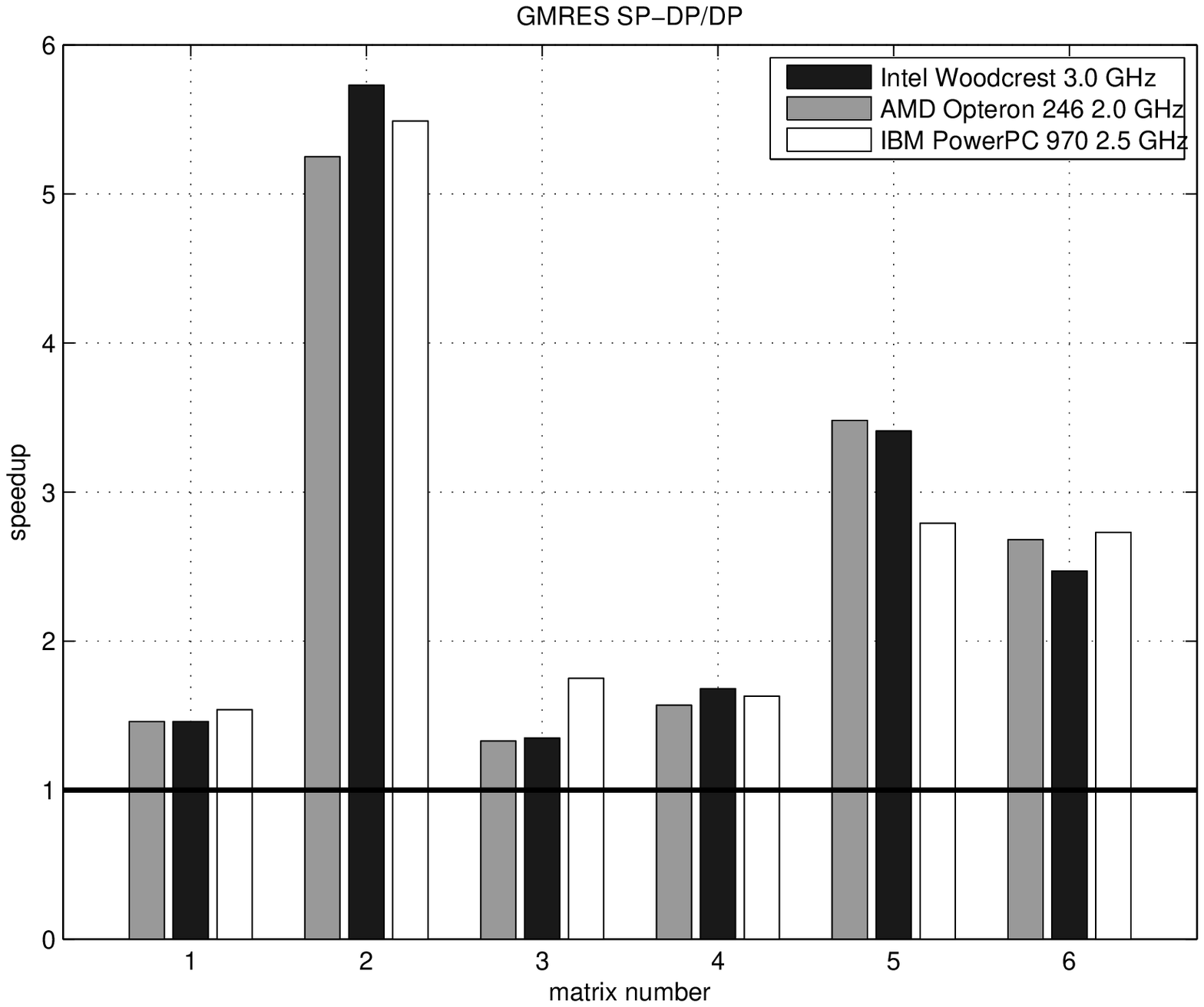}
\end{center}
\end{minipage}
  \caption{\label{fig:gmres}Mixed precision iterative refinement
    with FGMRES-GMRES$_{SP}$ from Algorithm~\ref{alg:pgmres}
    {\it vs } FGMRES-GMRES$_{DP}$ ({\it left}) and {\it vs} full double
    precision GMRES ({\it right}).}
\end{figure*}

Figure~\ref{fig:gmres}~({\it left}) shows the performance ratio of the mixed
precision inner-outer FGMRES-GMRES$_{SP}$ {\it vs.} the full double
precision inner-outer FGMRES-GMRES$_{DP}$. In other words, we compare two
inner-outer algorithms that are virtually the same. The only difference is that
their inner loop's incomplete solvers are performed in correspondingly single
and double precision arithmetic.

Figure~\ref{fig:gmres}~({\it right}) shows the performance ratio of the mixed
precision inner-outer FGMRES-GMRES$_{SP}$ {\it vs.} double precision GMRES.
This is an experiment that shows that inner-outer type iterative methods may be
very competitive compared to their original counterparts. For example, we
observe a speedup for matrix \#$2$ of up to $6$ which is mostly due to an
improved convergence of the inner-outer GMRES {\it vs.} standard GMRES. In
particular, about $3.5$ of the  $5.5$-fold speedup for matrix \# $2$ on the 
IBM PowerPC architecture is due to improved convergence, and the rest
$1.57$ speedup is due to single {\it vs} double precision arithmetic.
The restart values used for this computation are given in 
Table~\ref{tab:restarts}. The restart values $m_{in}$ and $m_{out}$
were manually tuned, $m$ was taken as $2 m_{out} + m_{in}$ 
in order to use the same amount of memory space for the two
different methods, or additionally increased when needed to improve
the reference GMRES solution times.

\begin{table}[!h]
  \centering
  \begin{tabular}{|cccc|}
\hline
matrix n. & $m_{in}$ & $m_{out}$ & $m$ \\
\hline
  1   &  30      &  20     &  150     \\
  2   &  20      &  10     &   40     \\
  3   & 100      &   9     &  300     \\
  4   &  10      &   4     &   18     \\
  5   &  20      &  20     &  300     \\
  6   &  20      &  10     &   50    \\
\hline
  \end{tabular}
  \caption{Restart values for the GMRES-based iterative solvers.}
  \label{tab:restarts}
\end{table}

\section{Numerical Remarks}
\label{sec:numeric}

Following the work of Skeel~\cite{Skeel_1980}, Higham~\cite{Higham_2002} gives error
bounds for the single and double precision, iterative refinement algorithm when
the entire algorithm is implemented with the same precision (single or double,
respectively).  Higham also gives error bounds in single precision arithmetic,
with refinement performed in double precision arithmetic~\cite{Higham_2002}.
The error analysis in double precision, for our mixed precision algorithm
(Algorithm~\ref{alg:dir_itref}), is given by Langou et
al.~\cite{Langou_2006_sc}.  Arioli and Duff~\cite{Arioli_Duff_2008} gives the
error analysis for a mixed precision algorithm based on a double precision
FGMRES preconditionned by a single precision LU factorization.  These errors
bounds explain that mixed precision iterative refinement will work as long as
the condition number of the coefficient matrix is smaller than the inverse of
the lower precision used.  For practical reasons, we need to resort to the
standard double precision solver in the cases when the condition number of the
coefficient matrix is larger than the inverse of the lower precision used.

\begin{figure}[!h]
\centering
  \includegraphics[width=0.8\textwidth]{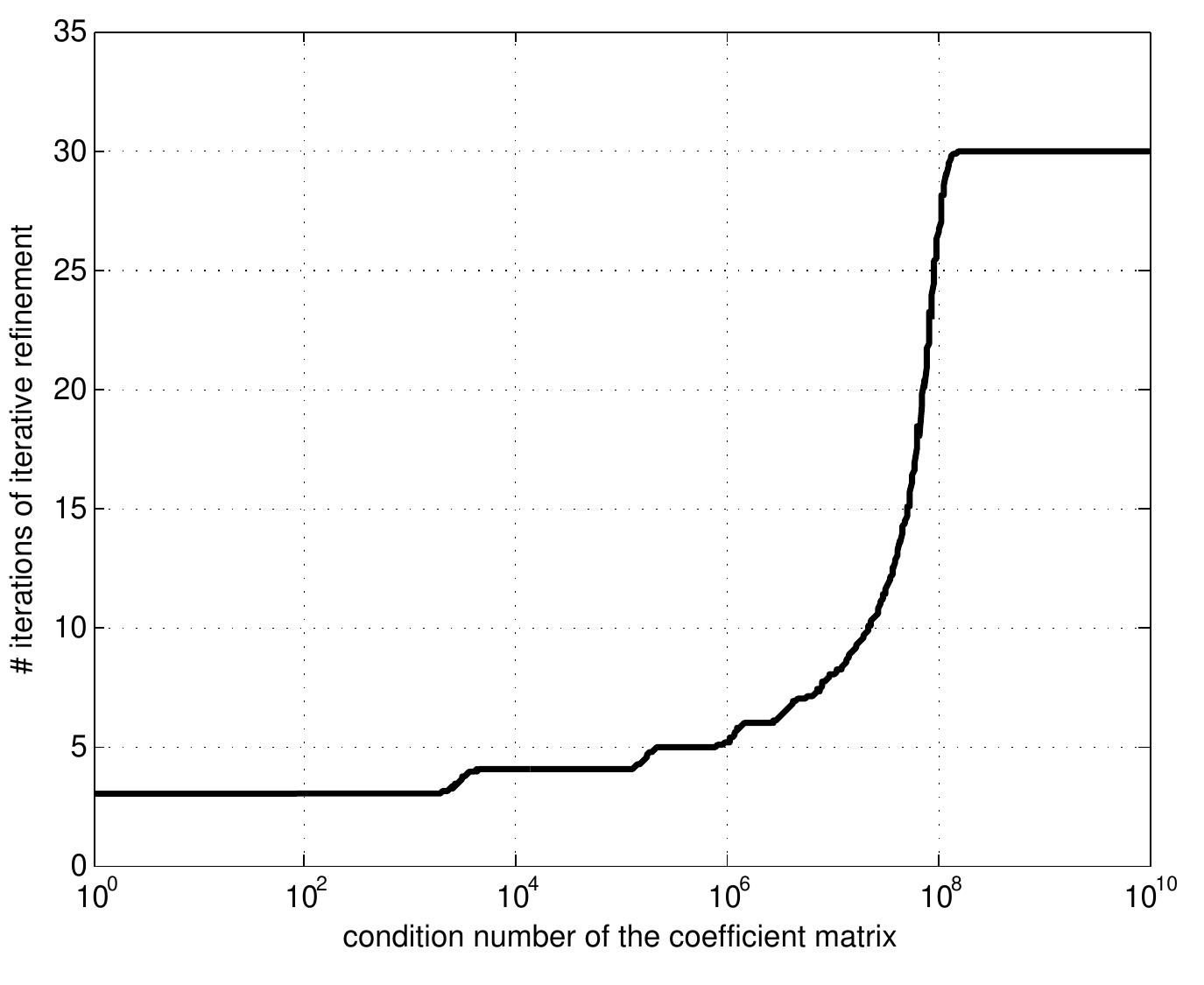}
  \caption{
Number of iterations needed for our mixed precision method to converge to an
accuracy better than the one of the associated double precision solve as a
function of the condition number of the coefficient matrix in the context of direct dense nonsymmetric solves.
}
  \label{fig:condition_number}
\end{figure}

In Figure~\ref{fig:condition_number}, we show the number of iterations
needed for our mixed precision method to converge to better accuracy than
the one of the associated double precision solve. The number of iterations is
shown as a function of the condition
number of the coefficient matrix ($\kappa$) in the context of direct dense nonsymmetric
solve. For each condition number, we have taken 200 random
matrices of size 200-by-200 with a prescribed condition number and we report the mean
number of iterations until convergence. The maximum number of iterations allowed
was set to 30 so that 30 means failure to converge (as opposed to convergence in 30
iterations). Datta~\cite{datta} has conjectured that the number of iterations necessary
for convergence was given by
$$ \left\lceil \frac{ \ln(\varepsilon_d) } { \ln(\varepsilon_d) + \ln(\kappa)}\right\rceil .$$
We can generalize this formula in the context of our mixed precision approach
$$ \left\lceil \frac{ \ln(\varepsilon_d) } { \ln(\varepsilon_s) + \ln(\kappa)}\right\rceil .$$

When $\kappa\varepsilon_s$ is above 1, then the formula is not valid anymore.
This is characterized in practice by an infinite number of iterations, i.e.
lack of convergence of the method.

\section{Extension to Quadruple Precision}

As an extension to this study, we present in this section results for
iterative refinement in quadruple precision on an Intel Xeon 3.2GHz. The
iterative refinement code computes a condition number estimate for input
matrices having random entries drawn from a uniform distribution. For quadruple
precision arithmetic, we use the reference BLAS compiled with
the Intel Fortran compiler \textit{ifort}  (with -O3 optimization flag
on) since we do not have an optimized BLAS in quadruple precision. The
version of the compiler is 8.1. Results are presented in Table~\ref{tab:quad-double--1}. The
obtained accuracy is between 10 and 32 for QGETRF and QDGETRF as expected. No
more than 3 steps of iterative refinement are needed. The speedup is
between 10 for a matrix of size 100 to close to 100 for a matrix of size
1000. In Table~\ref{tab:quad-double--2}, we give the time for the different kernels used in QGESV
and QDGESV. Interestingly enough the time for QDGESV is dominated by QGEMV
and not DGETRF! Recent research using related idea can be found in~\cite{gezh:03}.

\begin{table}[!h]
  \centering
  \begin{tabular}{|cccc|}
\hline
      & QGESV    & QDGESV   &         \\  
 n    & time (s) & time (s) & speedup \\
\hline
 100  & 0.29     & 0.03     &  9.5    \\
 200  & 2.27     & 0.10     & 20.9    \\
 300  & 7.61     & 0.24     & 30.5    \\
 400  & 17.81    & 0.44     & 40.4    \\
 500  & 34.71    & 0.69     & 49.7    \\
 600  & 60.11    & 1.01     & 59.0    \\
 700  & 94.95    & 1.38     & 68.7    \\
 800  & 141.75   & 1.83     & 77.3    \\
 900  & 201.81   & 2.33     & 86.3    \\
 1000 & 276.94   & 2.92     & 94.8    \\
\hline
  \end{tabular}
  \caption{Iterative Refinement in Quadruple Precision on a Intel Xeon 3.2GHz.}
  \label{tab:quad-double--1}
\end{table}

\begin{table}[!h]
  \centering
  \begin{tabular}{|cccc|}
\hline
driver name &  time (s) & kernel name &  time (s)   \\
\hline
QGESV   & 201.81  & QGETRF  & 201.1293 \\
        &         & QGETRS  &   0.6845 \\
\hline
QDGESV  &   2.33  & DGETRF  &   0.3200 \\
        &         & DGETRS  &   0.0127 \\
        &         & DLANGE  &   0.0042 \\
        &         & DGECON  &   0.0363 \\
        &         & QGEMV   &   1.5526 \\
        &         & ITERREF &   1.9258 \\
\hline
  \end{tabular}
  \caption{Time for the various Kernels in the Quadruple Accuracy Versions for n=900.}
  \label{tab:quad-double--2}
\end{table}

\section{Extension to Other Algorithms}

Mixed precision algorithms can easily provide substantial speedup for very
little code effort by mainly taking into account existing hardware properties.

We have shown how to derive mixed precision version of variety of algorithms
for solving general linear systems of equations.  Mixed precision iterative
refinement technique has also be used in the context of symmetric positive
definite systems~\cite{Kurzak_2007_cell_cholesky} using a Cholesky
factorization. In the context of overdetermined least squares problems, the
iterative refinement technique can be applied to the augmented system (where
both the solution and the residual are refined, as described
in~\cite{Demmel_2007}), to the QR factorization, to the semi-normal equations
or to the normal equations~\cite{Bjorck_1996}. Iterative refinement can also be
applied for eigenvalue computation~\cite{dongarra83_eig} and for singular value
computation~\cite{dongarra83_svd}.


We hope this manuscript will encourage scientists to extend this approach to
their own applications that do not necessarily originate from linear algebra.

\bibliographystyle{plain}
\bibliography{iter_ref}

\begin{thebibliography}{10}

\bibitem{amestoy2000}
P.~R. Amestoy, I.~S. Duff, and J.-Y. L'Excellent.
\newblock Multifrontal parallel distributed symmetric and unsymmetric solvers.
\newblock {\em Comput. Methods Appl. Mech. Eng.}, 184:501--520, 2000.

\bibitem{amestoy2001}
P.~R. Amestoy, I.~S. Duff, J.-Y. L'Excellent, and J.~Koster.
\newblock A fully asynchronous multifrontal solver using distributed dynamic
  scheduling.
\newblock {\em SIAM J. Matrix Analysis and Applications}, 23:15--41, 2001.

\bibitem{amestoy2006}
P.~R. Amestoy, A.~Guermouche, J.-Y. L'Excellent, and S.~Pralet.
\newblock Hybrid scheduling for the parallel solution of linear systems.
\newblock {\em Parallel Comput.}, 32:136--156, 2006.

\bibitem{lapack:99}
E.~Anderson, Z.~Bai, C.~Bischof, S.~Blackford, J.~Demmel, J.~Dongarra, J.~Du
  Croz, A.~Greenbaum, S.~Hammarling, A.~McKenney, and D.~Sorensen.
\newblock {\em {LAPACK} {U}sers' {G}uide}.
\newblock SIAM, Philadelphia, 3 edition, 1999.

\bibitem{Arioli_1989}
M.~Arioli, J.~W. Demmel, and I.~S. Duff.
\newblock Solving sparse linear systems with sparse backward error.
\newblock {\em SIAM J. Matrix Analysis and Applications}, 10(2):165--190, 1989.

\bibitem{Arioli_Duff_2008}
M.~Arioli and I.~S. Duff.
\newblock Using fgmres to obtain backward stability in mixed precision.
\newblock Technical Report RAL-TR-2008-006, Rutherford Appleton Laboratory,
  2008.

\bibitem{ashcraft1987}
C.~Ashcraft, R.~Grimes, J.~Lewis, B.~W. Peyton, and H.~Simon.
\newblock Progress in sparse matrix methods in large sparse linear systems on
  vector supercomputers.
\newblock {\em Intern.~J.~of Supercomputer Applications}, 1:10--30, 1987.

\bibitem{cg-vassilevski}
O.~Axelsson and P.~S. Vassilevski.
\newblock A black box generalized conjugate gradient solver with inner
  iterations and variable-step preconditioning.
\newblock {\em SIAM J. Matrix Anal. Appl.}, 12(4):625--644, 1991.

\bibitem{templates}
R.~Barrett, M.~Berry, T.~F. Chan, J.~Demmel, J.~M. Donato, J.~Dongarra,
  V.~Eijkhout, R.~Pozo, C.~Romine, and H.~V. der Vorst.
\newblock {\em Templates for the Solution of Linear Systems: Building Blocks
  for Iterative Methods.}
\newblock Philadalphia: Society for Industrial and Applied Mathematics., 1994.
\newblock Also available as postscript file at
  http://www.netlib.org/templates/Templates.html.

\bibitem{Bjorck_1996}
{\AA}.~{Bj\"orck}.
\newblock {\em Numerical Methods for Least Squares Problems}.
\newblock SIAM, 1996.

\bibitem{sparse_iterref}
A.~Buttari, J.~Dongarra, J.~Kurzak, P.~Luszczek, and S.~Tomov.
\newblock Using mixed precision for sparse matrix computations to enhance the
  performance while achieving 64-bit accuracy.
\newblock {\em ACM Trans. Math. Softw.}, 34(4):17, July 2008.
\newblock Article 17, 22 pages.

\bibitem{1297653}
A.~Buttari, J.~Dongarra, J.~Langou, J.~Langou, P.~Luszczek, and J.~Kurzak.
\newblock Mixed precision iterative refinement techniques for the solution of
  dense linear systems.
\newblock {\em Int. J. of High Performance Computing Applications},
  21(4):457--466, 2007.

\bibitem{datta}
B.~D. {Datta}.
\newblock {\em Numerical Linear Algebra and Applications}.
\newblock Brooks Cole Publishing Company, 1995.

\bibitem{davis1999}
T.~A. Davis.
\newblock A combined unifrontal/multifrontal method for unsymmetric sparse
  matrices.
\newblock {\em ACM Trans. Math. Softw.}, 25:1--19, 1999.

\bibitem{davis2004}
T.~A. Davis.
\newblock A column pre-ordering strategy for the unsymmetric-pattern
  multifrontal method.
\newblock {\em ACM Trans. Math. Softw.}, 30:196--199, 2004.

\bibitem{davis1997}
T.~A. Davis and I.~S. Duff.
\newblock An unsymmetric-pattern multifrontal method for sparse {LU}
  factorization.
\newblock {\em SIAM J. Matrix Analysis and Applications}, 18:140--158, 1997.

\bibitem{Demmel_1997}
J.~W. {Demmel}.
\newblock {\em Applied Numerical Linear Algebra}.
\newblock SIAM, 1997.

\bibitem{Demmel_2006}
J.~W. Demmel, Y.~Hida, W.~Kahan, X.~S. Li, S.~Mukherjee, and E.~J. Riedy.
\newblock Error bounds from extra-precise iterative refinement.
\newblock {\em ACM Trans. Math. Softw.}, 32(2):325--351, 2006.

\bibitem{Demmel_2007}
J.~W. Demmel, Y.~Hida, X.~S. Li, and E.~J. Riedy.
\newblock Extra-precise iterative refinement for overdetermined least squares
  problems.
\newblock Technical Report EECS-2007-77, UC Berkeley, 2007.
\newblock Also LAPACK Working Note 188.

\bibitem{dongarra83_svd}
J.~J. Dongarra.
\newblock Improving the accuracy of computed singular values.
\newblock {\em SIAM J. Scientific and Statistical Computing}, 4(4):712--719,
  December 1983.

\bibitem{blas05}
J.~J. Dongarra, J.~Du Croz, I.~S. Duff, and S.~Hammarling.
\newblock Algorithm 679: A set of level 3 basic linear algebra subprograms.
\newblock {\em ACM Trans. Math. Softw.}, 16:18--28, 1990.

\bibitem{blas04}
J.~J. Dongarra, J.~Du Croz, I.~S. Duff, and S.~Hammarling.
\newblock A set of level 3 basic linear algebra subprograms.
\newblock {\em ACM Trans. Math. Softw.}, 16:1--17, 1990.

\bibitem{blas03}
J.~J. Dongarra, J.~Du Croz, S.~Hammarling, and R.~J. Hanson.
\newblock Algorithm 656: An extended set of {FORTRAN} basic linear algebra
  subprograms.
\newblock {\em ACM Trans. Math. Softw.}, 14:18--32, 1988.

\bibitem{blas02}
J.~J. Dongarra, J.~Du Croz, S.~Hammarling, and R.~J. Hanson.
\newblock An extended set of {FORTRAN} basic linear algebra subprograms.
\newblock {\em ACM Trans. Math. Softw.}, 14:1--17, 1988.

\bibitem{dongarra83_eig}
J.~J. Dongarra, C.~B. Moler, and J.~H. Wilkinson.
\newblock Improving the accuracy of computed eigenvalues and eigenvectors.
\newblock {\em SIAM J. Numerical Analysis}, 20(1):23--45, February 1983.

\bibitem{duff1983}
I.~S. Duff and J.~K. Reid.
\newblock The multifrontal solution of indefinite sparse symmetric linear
  equations.
\newblock {\em ACM Trans. Math. Softw.}, 9(3):302--325, September 1983.

\bibitem{gezh:03}
K.~O. Geddes and W.~W. Zheng.
\newblock Exploiting fast hardware floating point in high precision
  computation.
\newblock In {\em Proceedings of the 2003 international symposium on Symbolic
  and algebraic computation. Philadelphia, PA, USA}, pages 111--118, 2003.

\bibitem{golub00inexact}
G.~H. Golub and Q.~Ye.
\newblock Inexact preconditioned conjugate gradient method with inner-outer
  iteration.
\newblock {\em SIAM J. Scientific Computing}, 21(4):1305--1320, 2000.

\bibitem{Higham_2002}
N.~J. {Higham}.
\newblock {\em Accuracy and Stability of Numerical Algorithms}.
\newblock SIAM, 2 edition, 2002.

\bibitem{Kurzak_2007_cell_cholesky}
J.~{Kurzak}, A.~{Buttari}, and J.~J. {Dongarra}.
\newblock Solving systems of linear equations on the {CELL} processor using
  {C}holesky factorization.
\newblock {\em IEEE Transactions on Parallel and Distributed Systems},
  19(9):1--11, September 2008.

\bibitem{Kurzak_2007_cell_linpack}
J.~{Kurzak} and J.~J. {Dongarra}.
\newblock Implementation of mixed precision in solving systems of linear
  equations on the {C}ell processor.
\newblock {\em Concurrency Computat.: Pract. Exper.}, 19(10):1371--1385, 2007.

\bibitem{Langou_2006_sc}
J.~{Langou}, J.~{Langou}, P.~{Luszczek}, J.~{Kurzak}, A.~{Buttari}, and J.~J.
  {Dongarra}.
\newblock Exploiting the performance of 32 bit floating point arithmetic in
  obtaining 64 bit accuracy.
\newblock In {\em Proceedings of the 2006 ACM/IEEE Conference on
  Supercomputing}, 2006.

\bibitem{blas01}
C.~L. Lawson, R.~J. Hanson, D.~Kincaid, and F.~T. Krogh.
\newblock Basic linear algebra subprograms for {FORTRAN} usage.
\newblock {\em ACM Trans. Math. Softw.}, 5:308--323, 1979.

\bibitem{Moler_1967_jacm}
C.~B. {Moler}.
\newblock Iterative refinement in floating point.
\newblock {\em J. ACM}, 14(2):316--321, 1967.

\bibitem{notay00flexible}
Y.~Notay.
\newblock Flexible conjugate gradients.
\newblock {\em SIAM J. Scientific Computing}, 22:1444--1460, 2000.

\bibitem{Oettli_1964}
W.~Oettli and W.~Prager.
\newblock Compatibility of approximate solution of linear equations with given
  error bounds for coefficients and right-hand sides.
\newblock {\em Numerische Mathematik}, 6:405--409, 1964.

\bibitem{saad91flexible}
Y.~Saad.
\newblock {A flexible inner-outer preconditioned {GMRES} algorithm}.
\newblock Technical Report 91-279, Department of Computer Science and
  Egineering, University of Minnesota, Minneapolis, Minnesota, 1991.

\bibitem{saad_book}
Y.~Saad.
\newblock {\em Iterative Methods for Sparse Linear Systems}.
\newblock Society for Industrial and Applied Mathematics, Philadelphia, PA,
  USA, 2003.

\bibitem{flexible-inner-outer}
V.~Simoncini and D.~B. Szyld.
\newblock Flexible inner-outer {K}rylov subspace methods.
\newblock {\em SIAM J. Numer. Anal.}, 40(6):2219--2239, 2003.

\bibitem{Skeel_1980}
R.~D. Skeel.
\newblock Iterative refinement implies numerical stability for {G}aussian
  elimination.
\newblock {\em Math. Comput.}, 35(151):817--832, 1980.

\bibitem{Stewart_1973}
G.~W. {Stewart}.
\newblock {\em Introduction to Matrix Computations}.
\newblock Academic Press, 1973.

\bibitem{Turner92}
K.~Turner and H.~F. Walker.
\newblock Efficient high accuracy solutions with {GMRES}(m).
\newblock {\em SIAM J. Sci. Stat. Comput.}, 13(3):815--825, 1992.

\bibitem{essg:03}
J.~van~den Eshof, G.~L.~G. Sleijpen, and M~.B. van Gijzen.
\newblock Relaxation strategies for nested {K}rylov methods.
\newblock {\em Journal of Computational and Applied Mathematics},
  177(2):347--365, 2005.

\bibitem{vandervorst91gmresr}
H.~A. van~der Vorst and C.~Vuik.
\newblock {GMRESR}: a family of nested {GMRES} methods.
\newblock {\em Numerical Linear Algebra with Applications}, 1(4):369--386,
  1994.

\bibitem{vuik-variable-GMRES}
C.~Vuik.
\newblock New insights in {GMRES}-like methods with variable preconditioners.
\newblock {\em J. Comput. Appl. Math.}, 61(2):189--204, 1995.

\bibitem{Wilkinson_1963}
J.~H. {Wilkinson}.
\newblock {\em Rounding Errors in Algebraic Processes}.
\newblock Prentice-Hall, 1963.

\bibitem{ypma:531}
T.~J. Ypma.
\newblock Historical development of the {N}ewton--{R}aphson method.
\newblock {\em SIAM Review}, 37(4):531--551, 1995.

\end{thebibliography}

\end{document}